\documentclass[prl,superscriptaddress,twocolumn]{revtex4-1}
\usepackage{latexsym}
\usepackage{amsfonts}
\usepackage{amssymb}
\usepackage{amsmath}
\usepackage{graphicx}

\begin{document}

\title{Spin-Torque Diode Measurements of MgO-Based Magnetic Tunnel Junctions with Asymmetric Electrodes}

\author{R. Matsumoto}
\email{rie.matsumoto@thalesgroup.com}
\affiliation{Unit\'e Mixte de Physique CNRS/Thales and Universit\'e Paris Sud 11, 1 ave A. Fresnel, 91767 Palaiseau, France}
\author{A. Chanthbouala}
\affiliation{Unit\'e Mixte de Physique CNRS/Thales and Universit\'e Paris Sud 11, 1 ave A. Fresnel, 91767 Palaiseau, France}
\author{J. Grollier}
\affiliation{Unit\'e Mixte de Physique CNRS/Thales and Universit\'e Paris Sud 11, 1 ave A. Fresnel, 91767 Palaiseau, France}
\author{V. Cros}
\affiliation{Unit\'e Mixte de Physique CNRS/Thales and Universit\'e Paris Sud 11, 1 ave A. Fresnel, 91767 Palaiseau, France}
\author{A. Fert}
\affiliation{Unit\'e Mixte de Physique CNRS/Thales and Universit\'e Paris Sud 11, 1 ave A. Fresnel, 91767 Palaiseau, France}
\author{K. Nishimura}
\affiliation{Process Development Center, Canon ANELVA Corporation, Kurigi 2-5-1, Asao, Kawasaki, Kanagawa 215-8550, Japan}
\author{Y. Nagamine}
\affiliation{Process Development Center, Canon ANELVA Corporation, Kurigi 2-5-1, Asao, Kawasaki, Kanagawa 215-8550, Japan}
\author{H. Maehara}
\affiliation{Process Development Center, Canon ANELVA Corporation, Kurigi 2-5-1, Asao, Kawasaki, Kanagawa 215-8550, Japan}
\author{K. Tsunekawa}
\affiliation{Process Development Center, Canon ANELVA Corporation, Kurigi 2-5-1, Asao, Kawasaki, Kanagawa 215-8550, Japan}
\author{A. Fukushima}
\affiliation{National Institute of Advanced Industrial Science and Technology (AIST) 1-1-1 Umezono, Tsukuba, Ibaraki 305-8568, Japan}
\author{S. Yuasa}
\affiliation{National Institute of Advanced Industrial Science and Technology (AIST) 1-1-1 Umezono, Tsukuba, Ibaraki 305-8568, Japan}

\begin{abstract}
We present a detailed study of the spin-torque diode effect in CoFeB/MgO/CoFe/NiFe magnetic tunnel junctions. From the evolution of the resonance frequency with magnetic field at different angles, we clearly identify the free-layer mode and find an excellent agreement with simulations by taking into account several terms for magnetic anisotropy. Moreover, we demonstrate the large contribution of the out-of-plane torque in our junctions with asymmetric electrodes compared to the in-plane torque. Consequently, we provide a way to enhance the sensitivity of these devices for the detection of microwave frequency.
\end{abstract}

\maketitle
Spin-transfer torque (STT) in MgO-based magnetic tunnel junctions (MTJs)\cite{Parkin1, Yuasa1} is under development for device applications such as STT random access memories (STT-RAM)\cite{Hosomi, Yuasa2}, domain-wall-motion MRAM,\cite{FukamiNEC} racetrack memory\cite{Parkin2} and spintronic memristors.\cite{SeagateIEEE, Andre} Recently, spin-torque diodes\cite{Tulapurkar} have attracted much attention because their sensitivity for the detection of microwave frequency may exceed that of semiconductor diodes.\cite{WangJAP, Ishibashi} In the spin diode effect, an applied rf current to the MTJ exerts an oscillating spin torque on the magnetization of the free layer, leading to excitation of the ferromagnetic resonance (FMR) mode. The dynamics of the free layer cause oscillations of the tunnel magnetoresistance (TMR). As a result, the oscillating resistance partially rectifies the rf current and dc voltage is obtained (\textit{V}$_{\rm diode}$). The spin diode effect depends on the relative amplitudes (\textit{a$_{J}$} and \textit{b$_{J}$}) of the classical in-plane torque (T$_{\rm IP}$) and the out-of-plane field-like torque (T$_{\rm OOP}$).\cite{Sankey, Kubota} Here, the torques are expressed as T$_{\rm IP}$ = -$\gamma($\textit{a$_{J}$}/\textit{M}$_{\rm s}$)\textbf{m}$\times$(\textbf{m}$\times$\textbf{M$_{\rm ref}$}) and T$_{\rm OOP}$ = -$\gamma$\textit{b$_{J}$}\textbf{m}$\times$\textbf{M$_{\rm ref}$} with \textbf{m} (\textit{M}$_{\rm s}$) being the magnetization vector (the saturation magnetization) of the free layer, \textbf{M$_{\rm ref}$} the magnetization vector of the reference layer, and $\gamma$ the gyromagnetic ratio. In spin diode spectra (\textit{V}$_{\rm diode}$ as a function of frequency (\textit{f})), the contribution of T$_{\rm IP}$ (resp. T$_{\rm OOP}$) results in a peak with a Lorentzian component (resp. an anti-Lorentzian component);
\setlength\arraycolsep{2pt}
\begin{eqnarray}
\mathbf{\textit{V}_{\rm diode}} & =  & \frac{A \left(f_1^2- f^2\right) + B f^2}{\left(f_1^2- f^2\right)^2 + \left(\Delta f \right)^2}, \label{Vdiode} \\
A &\propto& \gamma^2 H_{\rm d} \frac{\partial b_J}{\partial I} {\rm TMR \ sin^{2}}\theta, \label{A} \\
B &\propto& \gamma \Delta \frac{\partial a_J}{\partial I} {\rm TMR \ sin^{2}}\theta,\label{B} 
\end{eqnarray}
where \textit{A} and \textit{B} are the amplitudes of the anti-Lorentzian component and Lorentzian component, \textit{f}$_{\rm 1}$ is the resonance frequency, $\Delta$ is the peak linewidth, \textit{H}$_{\rm d}$ is the out-of-plane demagnetization field, and  $\theta$ is the relative angle between the free layer and the reference layer. Experimentally evaluated T$_{\rm OOP}$ was reported to reach over 25$\%$ of T$_{\rm IP}$ in conventional CoFeB/MgO/CoFeB MTJs with symmetric electrodes.\cite{Sankey, Kubota} However, in these MTJs, the dc bias dependence of T$_{\rm OOP}$ is quadratic and symmetric with respect to the polarity of bias, leading to \textit{A} = 0 at zero dc bias voltage. For MTJs with asymmetric electrodes, on the other hand, the bias dependence of T$_{\rm OOP}$ is expected to be asymmetric and linear at low bias,\cite{Xiao, Oh} leading to larger \textit{A} at zero dc bias voltage. In this study, we perform spin diode measurements of MgO-based MTJs with asymmetric electrodes. We also measure its dependence on magnitude and angle of the in-plane external magnetic field (\textit{H}$_{\rm ext}$) to identify the free-layer excitation modes.

Thin films of MTJs are deposited with a magnetron sputtering system (Canon ANELVA C-7100). The stacking structure (see Fig. 1(a)) is IrMn(7)/Co$_{70}$Fe$_{30}$(2.5)/Ru(0.9)/Co$_{60}$Fe$_{20}$B$_{20}$(3)/MgO tunnel barrier(1.1)/Co$_{70}$Fe$_{30}$(1)/Ni$_{83}$Fe$_{17}$(4)/capping layers (thickness in nm) deposited on thermally-oxidized Si substrate/buffer layers. Annealing treatment in a high-vacuum furnace at 330 $^{\circ}$C for 2 hours is then made under a 1 T magnetic field. These MTJ films are micro-processed into nano-pillars with an elliptic junction area of 70 ${\times}$ 270 nm$^{2}$. All the measurements presented here were done at room temperature. We first present in Fig. 1(b) the resistance versus magnetic field (\textit{R(H)}) curves obtained for the in-plane \textit{H}$_{\rm ext}$ applied along 0$^{\circ}$ and 90$^{\circ}$ (definition of the angle for \textit{H}$_{\rm ext}$ ($\phi$$_{\rm H}$) is schematically shown in an inset of Fig. 1(b)). The TMR ratio is 67.7$\%$ and the resistance is 162 $\Omega$ in the parallel magnetic state. The angular spin diode measurements presented in the following are performed with in-plane \textit{H}$_{\rm ext}$ at various $\phi$$_{\rm H}$. The injected microwave power is kept constant at -15 dBm, and no dc current is applied. 

\begin{figure}
	\includegraphics[width=7.5 cm]{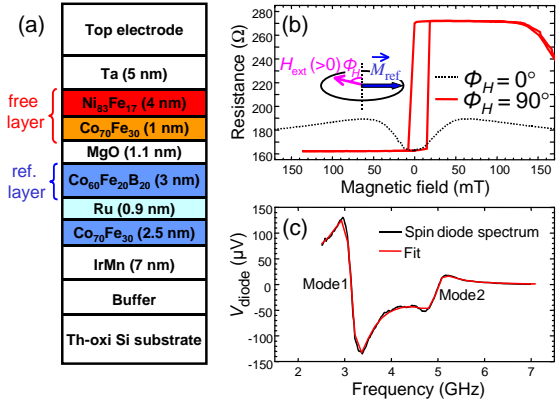}
\caption{(a) Sketch of the MgO-based magnetic tunnel junction (MTJ) stack. (b) Resistance versus magnetic field curves obtained for the in-plane magnetic field applied along 0$^{\circ}$(dotted line) and 90$^{\circ}$ (solid line). Inset in (b) is schematic explaining angle ($\phi_{\rm H}$) of in-plane magnetic field (\textit{H}$_{\rm ext}$) and magnetization direction of reference layer (\textit{M}$_{\rm ref}$). (c) Spin-torque diode signal (\textit{V}$_{\rm diode}$) versus frequency measured under \textit{H}$_{\rm ext}$ = +20 mT with {\textit{$\phi_{\rm H}$}} = 0$^{\circ}$ with fitted curve.}
\label{f1}
\end{figure}

As an example of spin diode measurements, in Fig. 1(c) we show the spectrum obtained at \textit{H}$_{\rm ext}$ = +20 mT with {\textit{$\phi_{\rm H}$}} = 0$^{\circ}$. In the MTJs of this study, the spin diode spectrum typically has two peaks that we label Mode 1 and Mode 2. As expected from Eq. (\ref{A}), in the case of the MTJs with asymmetric electrodes, the peaks have significant anti-Lorentzian component. The peaks are fitted well by Eq. (\ref{Vdiode}) (see Fig. 1(c)).

First, we focus on the property of Mode 1. The resonance frequency of Mode 1 versus \textit{H}$_{\rm ext}$ (\textit{f}$_{\rm 1}$(\textit{H})) at different angles is shown in Figs. 2(a), (c), and (e). For higher ${|}$\textit{H}$_{\rm ext}$${|}$ ${>}$ 500 Oe, \textit{f}$_{\rm 1}$(\textit{H}) increases with increasing \textit{H}$_{\rm ext}$ following Kittel formula.\cite{Kittel} Fitting \textit{f}$_{\rm 1}$(\textit{H}) for $\phi$$_{\rm H}$ = 0$^{\circ}$ with the Kittel formula gives \textit{H}$_{\rm d}$ = 1.1 T. For small \textit{H}$_{\rm ext}$ ${<}$ 500 Oe, \textit{f}$_{\rm 1}$(\textit{H}) exhibits an asymmetric behavior with respect to the polarity of \textit{H}$_{\rm ext}$, especially for \textit{H}$_{\rm ext}$ with $\phi$$_{\rm H}$ = 0$^{\circ}$ (Fig. 2(a)). This characteristic cannot be explained by macro-spin model taking into account only in-plane shape-anisotropy field (\textit{H}$_{\rm S.A.}$). The in-plane crystalline-anisotropy field (\textit{H}$_{\rm C.A.}$) due to Co$_{70}$Fe$_{30}$ in the free layer can be responsible for this asymmetric \textit{f}$_{\rm 1}$(\textit{H}). The CoFe layer is amorphous in the as-grown state but it crystallizes in a cubic-crystal texture by post-annealing\cite{Yuasa2} then leading to cubic anisotropy. It should be noted that its in-plane crystalline orientation can be arbitrary. 

\begin{figure}
	\includegraphics[width=7.5 cm]{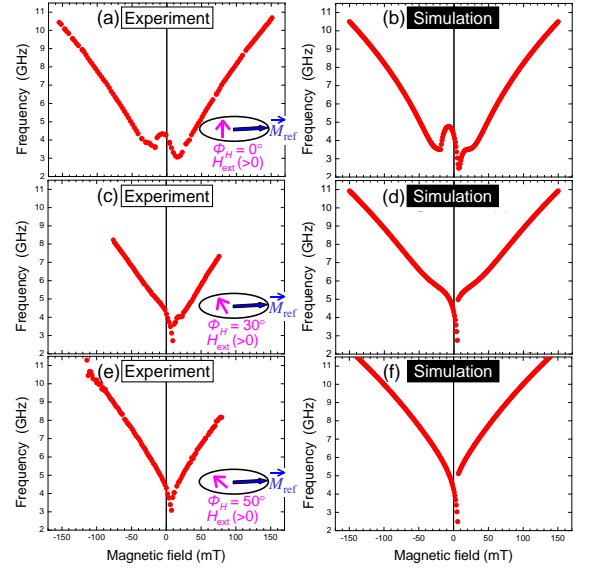}
\caption{Resonance frequency of free-layer mode versus external magnetic field (\textit{H}$_{\rm ext}$) with various angles (f) of (a)-(b) 0$^{\circ}$, (c)-(d) 30$^{\circ}$,  and (e)-(f) 50$^{\circ}$°. [(a), (c), (e)] Experimental results. Insets in (a), (b) and (c) are schematic explaining angle ($\phi$$_{\rm H}$) of in-plane magnetic field (\textit{H}$_{\rm ext}$) and magnetization direction of reference layer (\textit{M}$_{\rm ref}$). [(b), (d), (f)] Simulation results with consistent parameters: in-plane shape-anisotropy field (\textit{H}$_{\rm S.A.}$) of 20 mT, in-plane crystalline-anisotropy field (\textit{H}$_{\rm C.A.}$) of 10 mT, angle of crystalline-anisotropy field in CoFe layer ($\beta$) of 63$^{\circ}$, and \textit{H}$_{\rm d}$ = 1.1 T.}
\label{f2}
\end{figure}

To understand this asymmetric \textit{f}$_{\rm 1}$(\textit{H}), we analytically simulate the resonance frequency versus \textit{H}$_{\rm ext}$ of the free layer (\textit{f}$_{\rm free}$(\textit{H})) with macro-spin model taking into account not only \textit{H}$_{\rm S.A.}$ but also fourfold \textit{H}$_{\rm C.A.}$\cite{Goryunov} due to Co$_{70}$Fe$_{30}$ having an arbitrary crystalline orientation. First, the equilibrium angle of free-layer magnetization ($\phi$) under \textit{H}$_{\rm ext}$ with $\phi$$_{\rm H}$ is given by Eq. (\ref{Eq.Cond});
\begin{equation} \label{Eq.Cond}
\textit{H}_{\rm ext}{\rm sin}(\phi-\phi_{H})=\frac{1}{2}H_{\rm S.A.}{\rm sin}2\phi - \frac{1}{4}H_{\rm C.A.}{\rm cos}4(\phi-\beta).
\end{equation} 
Here, $\beta$ is the angle of crystalline easy axis in CoFe layer, defined with respect to the minor axis of the patterned ellipse.  
Then, \textit{f}$_{\rm free}$(\textit{H}) is given by Eq.(\ref{freq});
\begin{eqnarray} \label{freq}
\textit{f} & = & \frac{\gamma}{2\pi} \left[ H_{\rm d} \{H_{\rm ext} {\rm cos}(\phi-\phi_{H})\right. \nonumber\\
& &  \left.  -H_{\rm S.A.}{\rm cos}2\phi + H_{\rm C.A.}{\rm cos}4(\phi-\beta) \}\right]^\frac{1}{2}.
\end{eqnarray}
The simulation results are shown in Figs. 2 (b), (d), and (f). Here, we fix all parameters except for \textit{H}$_{\rm ext}$ and $\phi$$_{\rm H}$ : \textit{H}$_{\rm S.A.}$ = 20 mT, \textit{H}$_{\rm C.A.}$ = 10 mT, $\beta$ = 63$^{\circ}$, and \textit{H}$_{\rm d}$ = 1.1 T. The qualitative correspondence between the analytically simulated \textit{f}$_{\rm free}$(\textit{H}) and experimental \textit{f}$_{\rm 1}$(\textit{H}) indicates that Mode 1 corresponds to the free-layer mode, and that the contribution of \textit{H}$_{\rm C.A.}$ is not negligible. 

We also measure \textit{f}$_{\rm 1}$(\textit{H}) of other samples on the same lot. Although each sample exhibits the asymmetric \textit{f}$_{\rm 1}$(\textit{H}) with different characteristic at $\phi$$_{\rm H}$ = 0$^{\circ}$, each \textit{f}$_{\rm 1}$(\textit{H}) can be qualitatively explained by Eqs. (3) and (4) with various $\beta$ (0$^{\circ}$ - 90$^{\circ}$), similar values of \textit{H}$_{\rm S.A.}$ (20 - 40 mT) and \textit{H}$_{\rm C.A.}$ (15 - 20 mT), and the same \textit{H}$_{\rm d}$ = 1.1 T. This result also supports the measurable contribution of \textit{H}$_{\rm C.A.}$ To further obtain quantitative agreement between the simulated \textit{f}$_{\rm free}$(\textit{H}) and experimental \textit{f}$_{\rm 1}$(\textit{H}) especially in the low-field range, we need to take into account the coupling with Mode 2. However, the origin of Mode 2 is still under debate while such higher-order modes are often observed in MgO-based MTJs. Depending on the authors, the origin is attributed to edge mode, higher-order spin wave mode, or mode of CoFe/Ru/CoFeB synthetic antiferromagnetic layers.\cite{WangJAP, Helmer} To discuss the origin of Mode 2 is beyond the scope of this letter.

\begin{figure}
	\includegraphics[width=7.5 cm]{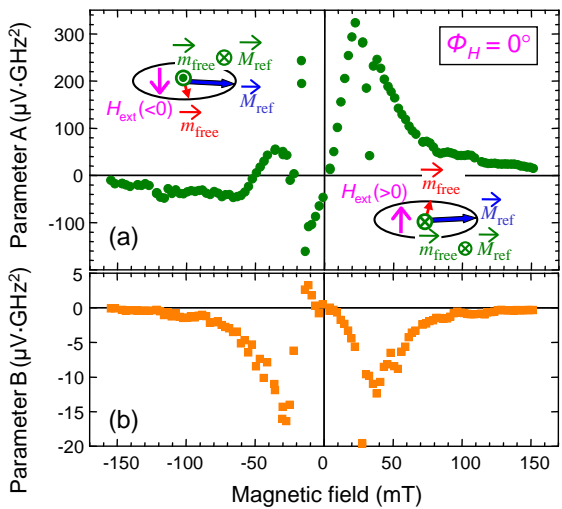}
\caption{Magnetic field dependence of (a) fitting parameter \textit{A} and (b) fitting parameter \textit{B} of Eq. (\ref{Vdiode}). Insets in (a) are schematic explaining angle ($\phi_{\rm H}$) of in-plane magnetic field (\textit{H}$_{\rm ext}$) and magnetization directions of free layer (\textit{m}$_{\rm free}$) and  reference layer (\textit{M}$_{\rm ref}$).}
\label{f3}
\end{figure}

The fitting parameters \textit{A} (anti-Lorentzian component) and \textit{B} (Lorentzian component) of Eq. (2) versus \textit{H}$_{\rm ext}$ with $\phi$$_{\rm H}$ = 0$^{\circ}$ that maximizes the spin diode effect are shown in Figs. 3 (a) and (b), respectively. First, the amplitude of the parameter \textit{A} is one order of magnitude larger than that of parameter \textit{B}. Indeed, the parameter \textit{A} is proportional to large $\gamma$\textit{H}$_{\rm d}$ whereas the parameter \textit{B} is proportional to small $\Delta$ $\approx$ $\gamma$$\alpha$\textit{H}$_{\rm d}$, where $\alpha$ is the Gilbert damping.\cite{Kubota, Sankey, WangPRB}. The linear bias dependence of T$_{\rm OOP}$ in the MTJs with asymmetric electrodes might open a way to enhance the diode sensitivity (\textit{V}$_{\rm diode}$ divided by injected rf power).  At \textit{H}$_{\rm ext}$ = 20 - 40 mT, the relative angle between the free layer and the reference layer ($\theta$) reaches its maximum. Then, it leads to maximum value of the parameter \textit{A} and \textit{B} for the chosen $\phi$$_{\rm H}$ = 0$^{\circ}$ because T$_{\rm IP}$ and T$_{\rm OOP}$ are proportional to sin $\theta$ (see Eqs. (\ref {A}) and (\ref {B})). 

We also check the diode sensitivity with a similar sample on the same lot. Here, we carefully measure the bias dc dependence of resistance and diode effect,\cite{Andre} and calibrate the impedance mismatch of the sample using the measured bias dependence of the resistance and backgrounds of the spin diode spectra.\cite{WangPRB} At \textit{H}$_{\rm ext}$ =  42 mT, we obtain a diode sensitivity of 100 mV/mW. This sensitivity is competitive compared to those of past studies although the TMR ratio in our study is about half.\cite{WangJAP, Ishibashi}

For higher ${|}$\textit{H}$_{\rm ext}$${|}$ ${>}$ 500 Oe, parameter \textit{A} changes its sign depending on the polarity of \textit{H}$_{\rm ext}$ while the sign of parameter \textit{B} is independent of the polarity of \textit{H}$_{\rm ext}$. This result agrees with the vectorial expression of the spin torques because the vector product in T$_{\rm OOP}$ changes polarity depending on the polarity of \textit{H}$_{\rm ext}$ as schematically shown in insets of Fig. 3, while the vector product in T$_{\rm IP}$ does not. Above-mentioned characteristics of \textit{H}$_{\rm ext}$ dependence of parameters \textit{A} and \textit{B} also support that Mode 1 corresponds to the free-layer mode.

In summary, spin diode measurements are performed in CoFeB/MgO/CoFe/NiFe MTJs having asymmetric electrodes without dc bias current. Their spin diode spectra exhibit peaks with strong anti-Lorentzian components originating from T$_{\rm OOP}$ in the asymmetric MTJs. Because of the significant contribution of T$_{\rm OOP}$, under an in-plane magnetic field of 42 mT, we obtain as large diode sensitivity as 100 mV/mW (after impedance matching correction). We also perform the spin diode measurements under various \textit{H}$_{\rm ext}$ with $\phi$$_{\rm H}$. By comparing experimental \textit{f}$_{\rm 1}$(\textit{H}) with the simulation results of \textit{f}$_{\rm free}$(\textit{H}),
we  can identify the free-layer excitation modes where the contribution of \textit{H}$_{\rm C.A.}$ is observed to be measurable as well as \textit{H}$_{\rm S.A.}$

\begin{acknowledgements}
This work was supported by the European Research Council (ERC Stg 2010 No. 259068) and JSPS Postdoctoral Fellowships for Research Abroad.
\end{acknowledgements}

\end{document}